\documentclass[runningheads,a4paper]{llncs}

\usepackage{amssymb}
\usepackage{graphicx}

\usepackage{url}
\newcommand{\keywords}[1]{\par\addvspace\baselineskip
\noindent\keywordname\enspace\ignorespaces#1}

\def\ins-figure#1#2#3{
\begin{figure}[t]
 \begin{center}
  \includegraphics[width=#3,keepaspectratio,clip]{#1.eps}
  \caption{#2}
  \label{fig:#1}
 \end{center}
\end{figure}
}

\def\fig#1{Fig.~\ref{fig:#1}}

\def\eq#1{Eq.~(\ref{eq:#1})}

\def\tab#1{Table~\ref{tab:#1}}

\begin{document}

\mainmatter  

\title{%
An Energy-efficient Time-domain Analog VLSI Neural Network Processor \\
Based on a Pulse-width Modulation Approach
}

\titlerunning{An Energy-efficient Time-domain Analog VLSI Neural Processor}

%
%
\author{Masatoshi Yamaguchi \and Goki Iwamoto %
\and Hakaru Tamukoh \and Takashi Morie}
\authorrunning{Masatoshi Yamaguchi \and Goki Iwamoto \and Hakaru Tamukoh \and Takashi Morie}

\institute{Graduate School of Life Science and Systems Engineering,\\
Kyushu Institute of Technology\\
2-4, Hibikino, Wakamatsu-ku, Kitakyushu, 808-0196 Japan\\
}

\maketitle

\begin{abstract}
A time-domain analog-weighted-sum calculation model based on
a pulse-width modulation (PWM) approach is proposed.
The proposed calculation model can be applied to any types of
 network structure including multi-layer feedforward networks.
We also propose very large-scale integrated (VLSI) circuits to implement the proposed model.
Unlike the conventional analog voltage or current mode circuits
 used in computing-in-memory circuits, 
our time-domain analog circuits use transient operation in charging/discharging processes to capacitors.
Since the circuits can be designed without operational amplifiers, 
they can be operated with extremely low power consumption.
However, they have to use very high-resistance devices, on the order of giga-ohms.
We designed a CMOS VLSI chip to verify weighted-sum operation based on the proposed model
with binary weights, which realizes the BinaryConnect model. 
In the chip, memory cells of static-random-access memory (SRAM)
are used for synaptic connection weights.
High-resistance operation was realized by 
using the subthreshold operation region of MOS transistors
unlike the ordinary computing-in-memory circuits.
The chip was designed and fabricated using a 250-nm fabrication technology.
Measurement results showed that energy efficiency for the weighted-sum calculation was 300~TOPS/W 
(Tera-Operations Per Second per Watt),
which is more than one order of magnitude higher than that in state-of-the-art 
digital AI processors, even though the minimum width of interconnection used in this chip 
was several times larger than that in such digital processors. 
If state-of-the-art VLSI technology is used to implement the proposed model, 
an energy efficiency of more than 1,000~TOPS/W will be possible. For practical applications, 
development of emerging analog memory devices such as ferroelectric-gate field effect 
transistors (FeFETs) is necessary.


\end{abstract}

\section{Introduction}

Artificial neural networks (ANNs), such as convolutional deep
neural networks (CNNs)~\cite{lecun98} and 
multi-layer perceptrons (MLPs)~\cite{ciresan10}, 
have shown excellent performance on various tasks 
including image recognition~\cite{ciresan10,krizhevsky12,farabet13,szegedy15,lecun15}.
However, computation in ANNs is very heavy, 
which leads to high power consumption in current digital computers 
and even in highly parallel coprocessors such as graphics processing units (GPUs). 
In order to implement ANNs at edge devices such as mobile phones and personal service robots, 
operation at very low power consumption is required.

In ANN models, weighted summation, or multiply-and-accumulate (MAC) operation, 
is an essential and heavy calculation task,
and dedicated complementary metal-oxide-semiconductor (CMOS)
very-large-scale integration (VLSI) processors have been 
developed to accomplish
it~\cite{sim16-isscc,moons17-isscc,shin17-isscc,khwa18-isscc,biswas18-isscc}.
As an implementation approach other than digital processors, 
use of analog operation in CMOS VLSI circuits is a promising method
for achieving extremely low-power consumption for such calculation 
tasks~\cite{fick17,lee16,miyashita17-jsc,mahmoodi18}.
In particular, computing-in-memory approaches,
which achieve weighted-sum calculation utilizing the circuit of 
static-random-access memory (SRAM), have been popular since around 2016~\cite{milojicic18}.

Although the calculation precision is limited 
due to the non-idealities of analog operation such as noise and device mismatches, 
neural network models and circuits can be designed to be robust to 
such non-idealities~\cite{morie94,indiveri02,guo17}.
On the other hand, ANN models with binarized weights or even with binarized inputs
have been proposed and their comparable performance has been demonstrated, mainly in
applications of image recognition~\cite{courbariaux15,hubara16}.
These models facilitate the development of energy-efficient hardware 
implementations~\cite{miyashita17-jsc}.

The time-domain analog weighted-sum calculation model was originally proposed based on 
mathematical spiking neuron models inspired by biological neuron behavior~\cite{maass97-nc,maass99-inbook}. 
We have simplified this calculation model under the assumption of operation in analog circuits
with transient states, and call its VLSI implementation approach 
``Time-domain Analog Computing with Transient states (TACT).''
In contrast to conventional weighted-sum operation in analog voltage or current modes, 
the TACT approach is suitable for operation with much lower power consumption in the CMOS VLSI implementation of ANNs.

We have already proposed a device and circuit that performs 
time-domain weighted-sum calculation~\cite{morie10-iscas,tohara16-apex,morie16-ieeenano}. 
The proposed circuit consists of plural input resistive elements 
and a capacitor (RC circuit), which can achieve extremely low-power operation.
The energy consumption could be lowered to the order of 1~fJ per operation, 
which is almost comparable to the calculation efficiency in the brain,
as long as weighted-sum operation is considered. 
We also proposed a circuit architecture to implement a weighted-sum calculation with different-signed weights
with two sets of RC circuits, one of which calculates positively weighted sums 
while the other calculates negatively weighted sums~\cite{wang16-iconip,wang18-arxiv}. 
Using a similar time-domain approach, a vector-by-matrix
multiplier using flash memory technology was proposed~\cite{bavandpour17-arxiv}.

Weighted-sum calculation circuits using pulse-width modulation (PWM) signals have previously been 
proposed~\cite{nagata98}.
In this paper, we reformulate the weighted-sum calculation model based on 
the time-domain analog computing approach using PWM signals, called the TACT-PWM approach,
and propose its applications to ANNs such as MLPs and CNNs 
with extremely high computing energy efficiency.
We also show the design and measurement results of an ANN VLSI chip 
fabricated using a 250-nm CMOS VLSI technology,
in which the calculation results by the proposed model are compared with 
the ordinary numerical calculation results and verify its very high computing efficiency.

\section{Time-domain weighted-sum calculation circuit model with PWM signals}

The basic circuit configuration based on the TACT-PWM approach is shown in \fig{PWM_SCS}. 
Corresponding to input signals $S_i \in \{0,1\}$ in the voltage domain, 
each switched-current source (SCS) outputs current $I_i$ when $S_i=1$. 
An SCS can be replaced by a resistor and a diode 
if the nonlinearity in charging characteristics can be ignored.
The total charge amount $Q$ stored at the node of capacitor $C$ charged by $N$ SCSs 
with inputs $S_i$, each of which has pulse width of $W_i$, is expressed by
\begin{equation}
Q = \sum_{i=1}^{N} W_{i} I_{i},
\end{equation}
where $Q$ can be considered as the weighted-sum calculation result 
with weight $I_i$ and input $W_i$.
The node voltage of $C$, $V_c$, is given by $V_c=Q/C$.
If $I_i \geq 0$, the energy consumption $E$ of this charging and discharging process is given by $E=CV_cV_{dd}$ 
($V_{dd}$ is a supply voltage of SCSs), 
where the energy for charging the input capacitance of SCSs is not included.

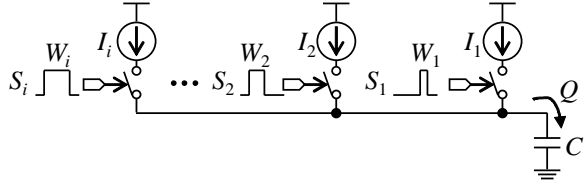
\ins-figure{PWM_SCS}{Weighted-sum calculation using current sources switched with PWM signals.}{80mm}

The weighted-sum calculation circuit and a timing diagram of its operation 
are shown in \fig{PWMmodel}.
Here, we consider this operation as a weighted-sum calculation with the same signed weighting.
The circuit consists of a weighted-sum calculation or MAC part and 
a voltage-pulse conversion (VPC) part.
The MAC part consists of SCSs corresponding with inputs, which 
is accompanied by parasitic wiring capacitance $C_d$.
The VPC part consists of an SCS, two switches, and a comparator with an input capacitance $C_n$.
Since the parasitic capacitances $C_d$ and $C_n$ are inevitably included in the circuit,
to minimize the energy consumption for the operation, the charged capacitance $C$,
which is equal to $C_d+C_n$, should be as small as possible.

The PWM inputs are given in the input period $T_{in}$; $\forall i, W_i \leq T_{in}$,
which is arbitrarily determined.
If the node voltage $V_c$ at the timing of the end of this input period is denoted by $V_{mac}$,
\begin{equation}
V_{mac} = \frac{Q}{C_d + C_n} = \frac{1}{C_d + C_n}\sum_{i=1}^{N} W_{i} I_{i}.
\end{equation}

In the VPC part, the output PWM signal $S_{out}$ with pulse width $W_{out}$ is generated 
during the output period $T_{out}$.
In this operation, capacitance $C$ is charged up by the SCS with current $I_n$.
To minimize the energy consumption in this operation, the VPC part 
can be separated from the MAC part by $S_n$, and only $C_n$ can be charged up
to the threshold voltage $V_\theta$ of the comparator. 
In this case, to meet the condition that 
$0 \leq W_{out} \leq T_{out}$, the current $I_n$ is given by
\begin{equation}
I_{n} = \frac{C_n V_\theta}{T_{out}},
\end{equation}
which means that the node voltage $V_n$ increases with the slope of $V_\theta/T_{out}$.
When $V_n > V_\theta$, the comparator output $S_{out}=1$, and after the end of output period
$V_n$ is reset by $S_{rst}$ at the resting state, which is usually zero.
Thus, the pulse width of the output signal as a result of weighted-sum calculation is given by
\begin{eqnarray}
W_{out} &=& \frac{V_{mac}}{V_{\theta}}T_{out} \\
         &=& \frac{T_{out}}{(C_d + C_n) V_\theta} \sum_{i=1}^{N} W_{i} I_{i},
\label{eq:tau}
\end{eqnarray}
where it is assumed that $0 \leq Q \leq (C_d + C_n)V_\theta$.

If the same input line structures are used regarding the positive and negative weights,
the denominator of \eq{tau} is common,
Thus, positive and negative weighted calculations are performed separately in the different lines, 
and by subtracting $W_{out}$ for negative weighing from that for the positive one,
the total calculation result is obtained as follows:
\begin{eqnarray}
W_{out}^+ - W_{out}^- &=& \frac{T_{out}}{(C_d + C_n)V_\theta} \left[\sum_{i=1}^{N+} W_{i}^+ I_{i}^+ 
                      - \sum_{i=1}^{N-} W_{i}^- I_{i}^-\right],\\
N &=& N^{+} + N^{-},
\end{eqnarray}
where $W_{out}^{\pm}$ are the pulse widths of output signals with positive and negative weighting, respectively.
Since the obtained result can be fed into the next circuit corresponding to the next layer of the network
via nonlinear transform operation, calculations for ANNs can be achieved.

The total energy consumption for the MAC calculation is expressed as follows:
\begin{eqnarray}
E_{cal} &=& E_{mac} + E_{vpc},\\
E_{mac} &=& C_d V_{mac}V_{dd} + \sum_{i=1}^{N} E_i,\\
E_{vpc} &=& C_n(V_{mac}+V_\theta)V_{dd} + E_n + \int_0^{T_{in}+T_{out}} P_{cmp}(t) dt,
\end{eqnarray}
where $E_{mac}$ and $E_{vpc}$ are the energy consumptions of the MAC and VPC parts, 
$E_i$ and $E_n$ are those for the switching of the SCS at each MAC part $i$ and
for the switching of the SCS at the VPC part, respectively, and $P_{cmp}(t)$ is the 
power consumption of the comparator.

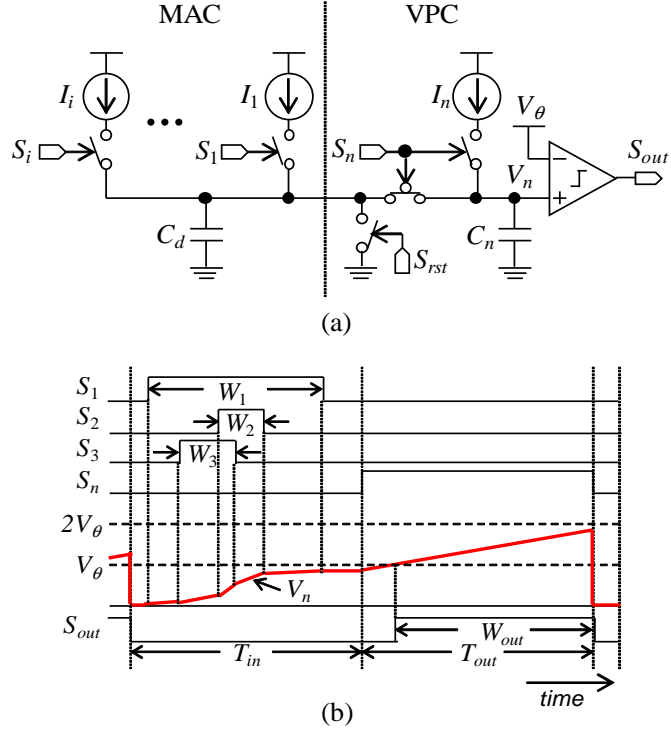
\ins-figure{PWMmodel}{Weighted-sum calculation circuit model with the same signed weighting: (a) circuit diagram and (b) timing diagram.}{90mm}

\section{CMOS BinaryConnect network circuit based on TACT-PWM approach}

On the basis of our TACT-PWM circuit approach, a CMOS circuit 
using an SRAM cell array structure is shown in \fig{PWMcircuit}(a).
This circuit implements a BinaryConnect neural network, which 
uses analog input values while weights are binary~\cite{courbariaux15}.

This circuit consists of a synapse part and a neuron part.
The synapse part consists of an SRAM cell array, and each synapse circuit operates as two MAC circuits.
Unlike the ordinary SRAM circuits proposed in the concept of computing-in-memory, 
our SRAM cell circuit outputs very low current on the order of nano-amperes 
to guarantee the time constant in the TACT approach~\cite{wang16-iconip,wang18-arxiv}, and therefore the 
p-type MOS field effect transistors (pMOSFETs) $M^{\pm}$
supply subthreshold currents to {\em dendrite} lines $D^{\pm}$ 
based on the input from {\em axon} lines $A_i$, 
where {\em axon} and {\em dendrite} are neuroscientific terms in the biological neuron.

In the neuron part, two VPC circuits perform positive and negative weighting calculations, respectively, 
and the subtraction result is fed into a rectified-linear-unit (ReLU) function circuit.
A detailed explanation follows.

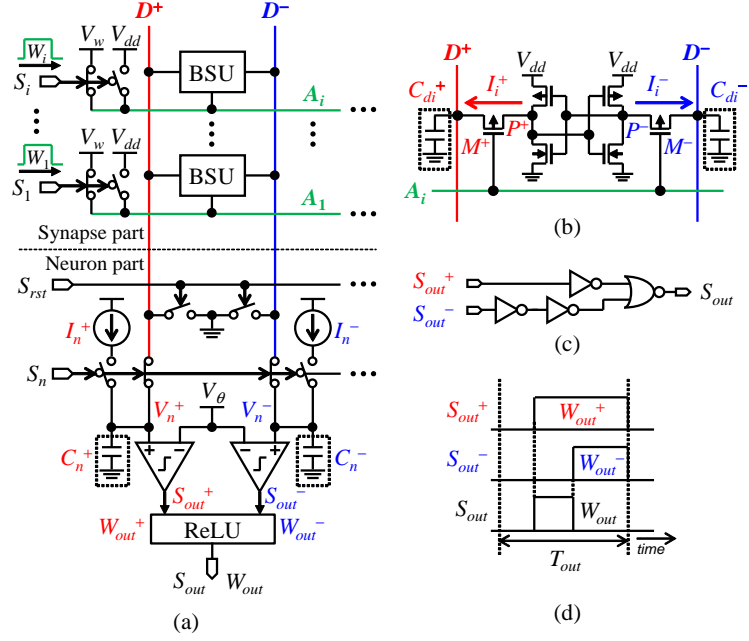
\ins-figure{PWMcircuit}{BinaryConnect neural network circuit based on TACT-PWM approach: 
(a) schematic diagram, (b) binary synapse unit (BSU) circuit, (c) ReLU function circuit,
and (d) timing diagram of the ReLU function circuit.}{100mm}

\subsection{Synapse part}

In the synapse part, each SRAM cell shown in \fig{PWMcircuit}(b), 
which is called here a binary synapse unit (BSU), performs binary weighting,
when receiving an input pulse $S_i$ as the gate voltage of the pMOSFET $M^{\pm}$ 
to make it operate in the subthreshold region.
To perform this operation, it is necessary that the SRAM cell be set at a 0 or 1 state 
based on the training result in a BinaryConnect network.

The BSU has three functions: one-bit memory, a switched current source, and a selector.
The one-bit memory function is achieved at the flip-flop, which stores the
binary weight $w_i \in \{+1,-1\}$ by setting voltages $V_{P}^{+}$ and $V_{P}^{-}$,
as follows:
\begin{eqnarray}
{w_{i}} &=&\{
\begin{array}{ll}
+1 & \mbox{if}\ \ (V_{P}^{+},V_{P}^{-})=(V_{dd},0)\\
-1 & \mbox{if}\ \ (V_{P}^{+},V_{P}^{-})=(0,V_{dd})
\end{array},
\end{eqnarray}
where $V_{dd}$ is the supply voltage.
The switched current source with a selector is realized by pMOSFETs $M^{\pm}$
that are connected to dendrite lines $D^{\pm}$, respectively. 
Since pMOSFETs $M^{\pm}$ operate in the subthreshold region, 
their drain currents $I_{i}^{\pm}$ are expressed as follows:
\begin{eqnarray}
I_{i}^{\pm} &\approx& I_0\exp(V_{P}^{\pm}-V_{Ai})\\
{V_{Ai}} &=&\left\{
\begin{array}{ll}
V_{dd} & \mbox{if}\ \ S_i=0\\
V_w & \mbox{if}\ \ S_i=1
\end{array}
\right\},
\end{eqnarray}
where $I_{0}$ is a constant, $V_{Ai}$ is the voltage of {\em axon} line $A_i$, 
and $V_w$ is the constant gate voltage for subthreshold operation.
For example, if synapse $i$ has positive weight ($w_i=1$) and $S_i=1$, then 
$(V_{P}^{+},V_{P}^{-})=(V_{dd},0)$, and $I_w^{+}\approx I_0\exp(V_{dd}-V_{w})$, and $I_w^{-}\approx0$.

\subsection{Neuron part}

In the neuron circuit, dendrite lines are initialized and reset at ground level by $S_{rst}$ before
inputting signals $S_i$ to the synapse part. 
Next, input PWM signals are given during input time period $T_{in}$, and
capacitance $C_{di}$ and $C_n$ are charged.
Then, dendrite lines are separated by neuron parts with $S_n$. 
At the same time, the current source $I_n$ is connected to capacitance $C_n$, 
and thus $C_n$ is charged. 
When the node voltage of $C_n$, $V_n^{\pm}$, reaches the threshold voltage of the comparator, 
the output signal $S_{out}^{\pm}$ is generated. A set of output signals $S_{out}^{\pm}$ are fed into 
the ReLU function circuit, which simply consists of logic circuits, 
as shown in \fig{PWMcircuit}(c), and
the output PWM signal is only generated when $W_{out}^{+} > W_{out}^{-}$, 
as shown in \fig{PWMcircuit}(d).

\section{VLSI chip design and measurement results}

Using TSMC 250~nm CMOS technology we designed and fabricated a CMOS VLSI chip of our neural network circuit 
with ten neurons each of which has 100 synapses.
The layout results and microphotographs are shown in \fig{layout}.

Measurement results of the input-output relationship in weighted-sum calculations operations
at one neuron with 100 synapses are shown in \fig{in_out_pulse}. 
As shown in \fig{in_out_pulse}(a), weighted-sum operation was approximately achieved
and sufficient linearity was obtained. From \fig{in_out_pulse}(b), the deviations in 
the time domain are $\pm 20$~ns, and this means that the precision of the calculation is about
$\pm 1$~\% because of the maximum pulse width being 2~$\mu$s.
However, an offset and scattering of weighting are clearly observed in \fig{in_out_pulse}(a). 
These nonidealities are due to variations in the threshold voltages of MOSFETs 
operating in the subthreshold region in BSUs.
Such variations can be compensated for by adjusting the threshold voltages 
if analog memory devices such as ferroelectric-gate FETs are used in BSUs.

Measurement results of the output pulse width as a function of 
weighted-sum calculation results followed by the ReLU function 
in one neuron with 100 synapses are shown in \fig{random_input}.
The average error was 1.5~\%, and the maximum error was about 8~\%.
This error can be decreased by adjusting the deviations of the threshold voltages 
of MOSFETs operating in the subthreshold region.

The measurement conditions and results for the power efficiency 
of the fabricated VLSI chip are shown in \tab{meas}. 
The power efficiency obtained from the measurement was 300~TOPS/W 
(Tera-Operations Per Second per Watt), which
is about 30 times higher than that of state-of-the-art digital AI processors, 
while the minimum feature size of the VLSI fabrication technology used was 
around 10 times larger than that in the digital AI processors.
Therefore, if we used the same VLSI fabrication technology as in the digital AI processors,
we could obtain a power efficiency of more than 1,000~TOPS/W or 1~POPS/W (Peta-OPS/W).

\begin{table}[tb]
\caption{Measurement conditions and results for power efficiency of the fabricated VLSI chip}
\begin{center}	
	\begin{tabular}{l|l}
\hline
	Number of synapses & 100 $\times$ 10  \\
	Operations per synapse & 2 (MAC) \\
	Number of neurons & 10 \\
	Input pulse width &  300 ns\\
	Output pulse width &  300 ns\\
	Supply voltage $V_{dd}$ &  1 V\\
	Threshold voltage $V_{\theta}$ &  0.2 V\\
	Operation freq. &  2.9E5 Hz\\
	Operations/sec &  5.9E8 OPS\\
\hline
	Power consumption & 1.9E-6 W \\
	Power efficiency & 3.0E14 OPS/W \\
\hline
	\end{tabular}
\end{center}
\label{tab:meas}
\end{table}

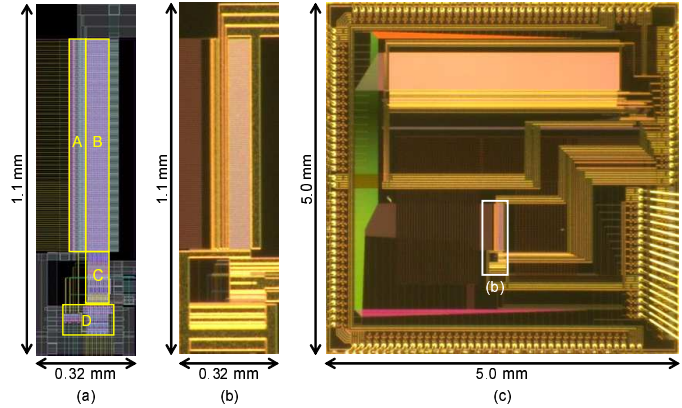
\ins-figure{layout}{VLSI layout results of a $100\times10$ BinaryConnect neural network: 
(a) layout result, (b) microphotograph of the circuit, and (c) chip microphotograph.
A: switch and buffer array for axon lines, 
B: BSU array, C: neuron array, 
and D: buffer array for dendrite lines. }{90mm}

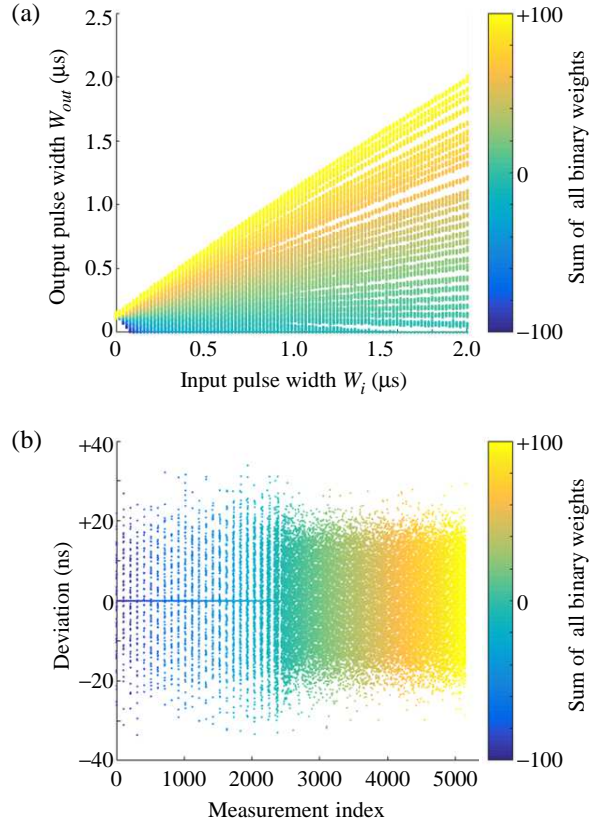
\ins-figure{in_out_pulse}{Measurement results of input-output characteristics:
(a) averaged output pulse width and (b) deviation.}{80mm}

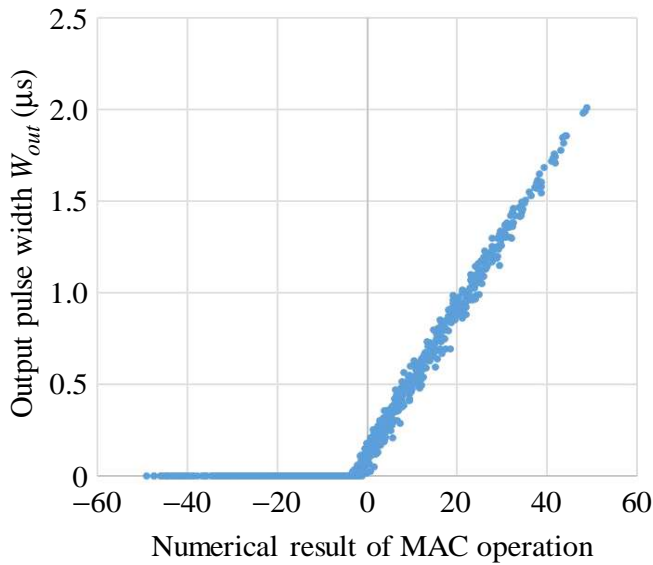
\ins-figure{random_input}{Measurement results of output pulse widths for the combination of random weights and inputs.
Timing jitters were decreased by averaging output signals for 50 measurement results.
The horizontal axis shows numerical calculation values of 
$\sum_{i=1}^{N=50} w_{i} \cdot W_{i}/T_i$,
where $w_i \in \{+1,-1\}$ and $0 \leq W_i/T_{in} \leq 1$.}{90mm}

\section{Conclusions}

In this paper, we proposed a time-domain weighted-sum calculation model based
on the TACT-PWM approach with an activation function of ReLU.
We also proposed VLSI circuits based on the TACT approach
to implement a calculation model with extremely low energy consumption.
A high energy efficiency of 300~TOPS/W was achieved by the
fabricated CMOS VLSI circuit with binary weights using 250-nm CMOS VLSI technology. 
If we use a more advanced VLSI fabrication technology, 
which achieves lower parasitic capacitance,
the energy efficiency will be further much improved to over 1,000~TOPS/W.

However, the fabricated circuit had insufficient calculation precision, 
which is mainly due to the characteristic variations of subthreshold 
operation in MOSFETs.
To improve the calculation precision and compensate for such variations, it is necessary 
to introduce analog memory devices.

As for the neuron parts, the measurement results of the fabricated VLSI chip 
suggest that the energy consumption of this part is comparable to 
that of the whole synapse part with 100 inputs. 
Therefore, it is also necessary to redesign a comparator circuit
with much lower power consumption to improve the energy efficiency 
of the whole calculation circuit.

\subsubsection*{Acknowledgments.}

This work was supported by JSPS KAKENHI Grant Nos. 22240022 and 15H01706. 
Part of the work was carried out under a project commissioned by the New Energy and Industrial Technology Development Organization (NEDO),
and the Collaborative Research Project of the Institute of Fluid Science, Tohoku University.
The circuit design was supported by VLSI Design and Education Center (VDEC),
the University of Tokyo in collaboration with Cadence Design Systems, Inc.,
Mentor Graphics, Inc., and Synopsys, Inc.


\bibliographystyle{splncs03}
\bibliography{/_Work/my-nn,/_Work/my-group}

\end{document}